\pdfoutput=1
\documentclass[%
 reprint,
superscriptaddress,
preprintnumbers,
nofootinbib,
 amsmath,amssymb,
 aps,
]{revtex4-1}
\usepackage{graphicx}
\usepackage{dcolumn}

\newcommand{\be}{\begin{equation}}
\newcommand{\ee}{\end{equation}}
\newcommand{\ba}{\begin{eqnarray}}
\newcommand{\ea}{\end{eqnarray}}

\renewcommand{\l}{\left(}
\renewcommand{\r}{\right)}

\begin{document}

\preprint{INR-TH/2014-027}

\title{\bf Collider signatures of Hylogenesis}

\author{S.\,V.\,Demidov}
 \email{demidov@ms2.inr.ac.ru}
\affiliation{Institute for Nuclear Research of the Russian Academy of Sciences,
  Moscow 117312, Russia}%
\author{D.\,S.\,Gorbunov}
 \email{gorby@ms2.inr.ac.ru}
\affiliation{Institute for Nuclear Research of the Russian Academy of Sciences,
  Moscow 117312, Russia}%
\affiliation{Moscow Institute of Physics and Technology, 
  Dolgoprudny 141700, Russia}%
\author{D.\,V.\,Kirpichnikov}
 \email{kirpich@ms2.inr.ac.ru}
\affiliation{Institute for Nuclear Research of the Russian Academy of Sciences,
  Moscow 117312, Russia}%


\begin{abstract}
We consider collider signatures of the hylogenesis --- a variant of
antibaryonic dark matter model.  We obtain bounds on the model
parameters from results of the first LHC run. Also we suggest
several new channels relevant for
probing the antibaryonic dark matter at LHC.
\end{abstract}

\maketitle

\section{Introduction}

The concept of particle dark matter (DM) is the great mystery in
physics\,\cite{Trimble:1987ee}. In spite of the fact that DM
constitutes a sizable fraction $\Omega_{\text{DM}}$ of the present
Universe energy density and has amount almost five times larger than
that of baryons $\Omega_{\text{B}}$\,\cite{Hinshaw:2012aka} particle
physics experiments have so far failed to obtain any convincing signal
from dark matter sector.

Existence of dark matter is one of the main phenomenological reasons
to believe that Standard Model of particle physics should be replaced
by a new theory.  One of the explanations for DM is existence of a
cosmologically stable weakly interacting massive particles
(WIMPs). Being thermalized in the primordial plasma at the Hot stage
of the Universe evolution WIMPs later become nonrelativistic and
freeze-out through annihilation into SM particles. Then their relic
abundance is determined by the corresponding annihilation cross
section. Such particles can naturally appear in some SM extensions
like supersymmetry or extra dimensions. However, in such theories in
general the fact that $\Omega_{\text{DM}}$ is of the same order as
$\Omega_{\text{B}}$ is accepted as a coincidence (``the WIMP
miracle'').

However, the latter coincidence may be a hint at the common origin of
baryonic and dark matter. Moreover, this may be related to another
important problem of the Standard Model - asymmetry between number of
particles and antiparticles in our visible Universe. This line of
reasoning resulted in construction of asymmetric dark matter models
where dark matter candidates have a net (anti)-baryonic charge (see,
e.g.\,\cite{Cohen:2010kn,Hooper:2004dc,Kitano:2004sv,Agashe:2004bm,Farrar:2005zd,Shelton:2010ta,Cheung:2011if}
and Ref.\,\cite{Petraki:2013wwa} for a review).

An attractive effective model of asymmetric dark
matter---hylogenesis---has been proposed in\,\cite{Davoudiasl:2010am}
(a particular high-energy completion of the hylogenesis scenario in
the framework of supersymmetric models is presented
in\,\cite{Blinov:2012hq}). This model exhibits several distinct
phenomenological signatures: {\it (i)} the induced nucleon 
decay (IND)\,\cite{Davoudiasl:2010am,Davoudiasl:2011fj,Huang:2013xfa}; {\it
(ii)} the direct production of dark matter particles in particle
collisions, in particular, at LHC~\cite{Davoudiasl:2011fj}. The idea
of collider searches for dark
matter~\cite{Goodman:2010ku,Fox:2011pm,Rajaraman:2011wf} has 
recently got much attention especially with on-going LHC
experiments. This kind of searches is especially important for the
class of models with asymmetric dark matter (see
e.g.~\cite{Kim:2013ivd} for a review) where other standard searches
based on direct detection or indirect searches for annihilation
products are not (so) effective. 

The goal of this paper is to get bounds on parameters of the
hylogenesis model from searches for processes with a monojet and
missing energy in the 1st LHC run data and to estimate the signal for
future LHC runs. This is the common signature for all dark matter
models including WIMPs, though in hylogenesis the jet is originated
from the new physics interaction rather than incoming quark or gluon
bremsstrahlung. Apart from this we discuss also a way of
discriminating the hylogenesis mechanism from other dark matter models
by looking for the process with four jets, three of which emerge from
a heavy state decay.

The paper is organized as follows. In Sec.\,\ref{sec:1} we describe
the main features of hylogenesis mechanism. In Sec.\,\ref{sec:2} we
discuss bounds on its parameters obtained from results of searches for
single jet and missing energy in the 1st LHC run. In Sec.\,\ref{sec:3}
we estimate sensitivity of the future LHC runs and suggest other
possible collider signatures of the hylogenesis. The last Section
contains discussion and conclusions.


\section{The model}
\label{sec:1}
In the hylogenesis model\,\cite{Davoudiasl:2010am} the asymmetric dark
matter consists of two components: fermion $Y$ and scalar $\Phi$.
Each carries nonzero baryonic charge such that $B_{Y} + B_{\Phi}
= -1$. These fields couple to visible matter via 
``neutron''-like portal as  
\be
\label{eq:1}
{\cal L}\! = \!-\!\!\sum_{a=1,2}\!\!\frac{\lambda^{ijk}_a}{M^2}\bar{X}_aP_Rd^i
\bar{u}^{jC}P_Rd^{k} + \zeta_a\bar{X}_aY^C\Phi^* + {\rm h.c.}, 
\ee
where $X_a, a=1,2$ are heavy fermionic mediators, indices $i,j,k$
label generations and superscript $C$ denotes 
charge conjugation. In the first term of\,\eqref{eq:1} we implicitly
assume that the color indices are convoluted to form a color singlet. 
Masses of $X_{1,2}$ are supposed to be at TeV
scale and obey $m_{X_2}>m_{X_1}$. Nonrenormalizable
interaction in 
Eq.\,\eqref{eq:1} is suppressed by a high energy scale $M$ where the
model must be UV completed to become renormalizable. 
Baryon asymmetry is generated by CP-asymmetric decays of nonthermal
population of $X_1$ and $\bar{X}_1$. The rate of the 
dominant decay mode
$X_1\to \bar{Y}\Phi^*$ is
\be
\label{A}
\Gamma(X_1\to \bar{Y}\Phi^*) = \frac{|\zeta_1|^2\,m_{X_{1}}}{16\,\pi}\,.
\ee
These heavy fermions can decay also as $X_1\to udd$
and $\bar{X}_1\to\bar{u}\bar{d}\bar{d}$ (for coupling to the first
quark generation) with tree level decay rate 
\be
\label{B}
\Gamma(X_1\to udd)
= \frac{3|\lambda_1|^2\,m_{X_1}^5}{1024\,\pi^3\,M^4}. 
\ee
The baryon asymmetry in the visible sector is generated at one-loop
level by the latter process (see Ref.\,\cite{Davoudiasl:2010am} for
details), which yields for the microscopic
asymmetry (asymmetry per one decay of $X_1, \bar{X}_1$ pair)    
\be
\begin{split}
\delta &\equiv \frac{1}{2\Gamma_{X_1}}\left(\Gamma(X_1\to udd) -
\Gamma(\bar{X}_1\to\bar{u}\bar{d}\bar{d})\right) \\&\approx 
\frac{m_{X_1}^5{\rm
    Im}\left[\lambda^*_1\lambda_2\zeta_1\zeta^*_2\right]} 
     {256\pi^3|\zeta_1|^2M^4m_{X_2}}
\\&\sim 10^{-4}\times \frac{{\rm
    Im}\left[\lambda^*_1\lambda_2\zeta_1\zeta^*_2\right]}{|\zeta_1|^2}\, 
\l\frac{m_{X_1}}{M}\r^4\,\l\frac{m_{X_1}}{m_{X_2}}\r
\end{split}
\label{BAU}
\ee
assuming $m_{X_2}\gg m_{X_1}$. 

Large enough asymmetry requires
that the scale $M$ should not significantly exceed the masses of
$X_{1,2}$. Indeed, if generated at hot stage without any suppression
and subsequent washing out the macroscopic asymmetry $\Delta_B\equiv
(n_B-n_{{\bar B}})/s$ is estimated as $\Delta_B\sim\delta/g_*$, 
where $s$ and $g_*$ are the entropy density and number of relativistic
degrees of freedom in plasma at that epoch. Macroscopic asymmetry
remains constant later, while the Universe expands, so one has $\Delta_B\simeq
10^{-10}$\,\cite{PDG}, hence $\delta \sim 10^{-8}$ for anticipated
$g_*\sim 10^2$. If the macroscopic asymmetry gets suppressed, than
successful baryogenesis asks for larger $\delta$. Introducing 
small parameter $\epsilon$, so that $m_{X_1}\sim \epsilon m_{X_2}$ and 
 $m_{X_2}\sim \epsilon M$ one observes from \eqref{BAU}, that even
 $\epsilon<0.3$ is unacceptable without a hierarchy in coupling
 constants, $\zeta_a\sim\lambda_a\sim 1$. With strong hierarchy 
$\zeta_1\ll\zeta_2$
 smaller values,  $\epsilon\ll 0.3$, become allowed.

Both dark matter particles $Y$ and $\Phi$ 
are supposed to be at GeV scale and their
stability is a consequence of baryon number conservation and
kinematical constraints on their masses: $|m_{Y}-m_{\Phi}| < m_p +
m_e$. In this mechanism the net baryonic charge of the Universe is
zero and it is supposed to be perturbatively 
conserved during the late-time evolution of the
Universe. As a consequence relic number densities of baryons and dark
matter particles are related as $n_{Y}=n_{\Phi}=n_B$ and using
cosmological data this gives the following mass range for 
dark matter particles: $1.7$~GeV~$\lesssim
m_{Y},m_{\Phi}\lesssim$~2.9~GeV. 

Light dark matter particles can rescatter into quarks, $Y\Phi\to qqq$, 
and wash out the generated in the decays of $X$ asymmetry. This process is
ineffective if reheating temperature is sufficiently
low \cite{Davoudiasl:2010am} 
\be
\label{BBN}
\frac{T_{rh}}{2\,\text{GeV}}\lesssim 
\l \sum_{a,b} \lambda_a\lambda_b^* \zeta^*_a\zeta_b 
\frac{\text{TeV}^6}{M^4\, m_{X_a}\,m_{X_b}}
\r^{-1/5},
\ee 
yet it must exceed few MeV required by the standard Big Band
Nucleosynthesis\,\cite{PDG}. 
Note that mild hierarchy $\epsilon\sim
0.3$ is allowed by \eqref{BBN}. Moreover, the new particles may
be sufficiently below TeV scale. Say, with
$\lambda_1\sim\lambda_2\sim\zeta_2\sim 1$ and $\zeta_1\sim 10^{-5}$
and $\epsilon\sim 0.1$ the process \eqref{A} still dominates
over \eqref{B}, while asymmetry \eqref{BAU} is at acceptable level,  
$\delta \sim 10^{-8}$ and reheating temperature \eqref{BBN} is above
MeV scale for $M=3$\,TeV, while the mass spectrum is $m_{X_1}\sim
30$\,GeV, $m_{X_2}\sim 300$\,GeV.

Hylogenesis mechanism can work for different combinations 
of flavors in the interaction\,\eqref{eq:1}, each is 
characterized by coupling $\lambda_{a}^{ijk}$. Hereafter we analyze
the simplest cases where only one of these constants is non-zero and
contributes to both the baryon asymmetry 
generation and the production
of dark matter particles in proton-proton collisions. 
We consider four cases (models) with different types of interaction 
\begin{align}
\label{first-a}
{\cal O}^{dud} & = -\frac{\lambda_a^{dud}}{M^2}(\bar{X}_a P_R
d)(\bar{u}^{C} P_R d), \\
\label{first-b}
{\cal O}^{dus} & = -\frac{\lambda_a^{dus}}{M^2}(\bar{X}_a P_R
d)(\bar{u}^{C} P_R s), \\
\label{second-a}
{\cal O}^{dub} & = -\frac{\lambda_a^{dub}}{M^2}(\bar{X}_a P_R
d)(\bar{u}^{C} P_R b), \\ 
\label{second-b}
{\cal O}^{dtd} & = -\frac{\lambda_a^{dtd}}{M^2}(\bar{X}_a P_R
d)(\bar{t}^{C} P_R d),
\end{align}
where again convolution of color indices for quarks with $SU(3)$
antisymmetric tensor is implicitly assumed. Sure enough one can write
down also operators with different permutations of quarks
fields. Collider experiments are sensitive to all types of
interactions presented above but the operators in\,\eqref{first-a},
\eqref{first-b}   
also provide with IND signatures
discussed in~\cite{Davoudiasl:2011fj,Huang:2013xfa} and can be probed 
in this way. As to operators\,\eqref{second-a}, \eqref{second-b}   
collider searches is the only way to probe this
scenario, since contributions from  \eqref{second-a}, \eqref{second-b}
to proton decay are suppressed due to absence of 
the sea heavy quarks in proton. 

TeV scale for $X_{1,2}$ was adopted 
in \cite{Davoudiasl:2010am} because for smaller masses the proton
lifetime starts to exceed the typical upper limits from
Super-Kamiokande (see e.g. \cite{Kobayashi:2005pe,Abe:2014mwa}). Though 
the kinematics of IND (due to dark matter inelastic scattering) 
is different, so that Super-Kamiokande bounds are not directly applicable,
light $X_{1,2}$ are expected to be disfavored from these searches. If
operators \eqref{first-a}, \eqref{first-b} are suppressed (absent),
while \eqref{second-a}, \eqref{second-b} are at work, the proton
limits are significantly weaker and lower masses are acceptable even
without strong hierarchy in coupling constants provided 
$\delta\gtrsim 10^{-8}$. With a hierarchy in coupling constants
(e.g. like one suggested above) the masses may be significantly
lower.

\section{LHC bounds}
\label{sec:2}
Here we discuss bounds on the model parameters 
from results of LHC searches for monojet and missing 
energy
signature~\cite{Chatrchyan:2011nd,Aad:2011xw,Khachatryan:2014rra}. 
Sensitivity of the collider (Tevatron and LHC) searches to hylogenesis
model has been discussed in~\cite{Davoudiasl:2011fj}. Below we 
consider the direct production of $X_{1,2}$ in association
with a single jet. Since $X_{1,2}$ decay dominantly into dark matter particles
$Y$ and $\Phi$ the corresponding events exhibit missing energy
signature. 
In our calculations we limit ourselves to the case when narrow width
approximation is valid.  
Corresponding tree level
differential cross sections for different operators 
in~\eqref{first-a}-\eqref{second-b} look as follows  
\begin{itemize}
\item Operator ${\cal O}^{dud}$. There are two main
  subprocesses contributing to  
  reaction $pp\to jet + E_T^{miss}$: subprocess $d(p_1)d(p_2)\to
  \bar{u}(k_1)X_{a}(k_2)$ with differential cross section
\be
\label{operator-1}
\frac{d\hat{\sigma}}{dt} = 
\frac{|\lambda^{dud}_a|^2}{96\pi s^2M^4}\!\left[4t^2 +
4t(\!s-m_{X_a}^2) + s(s-m_{X_a}^2)\right]
\ee
and subprocess $d(p_1)u(p_2)\to \bar{d}(k_1)X_{a}(k_2)$ with
\be
\frac{d\hat{\sigma}}{dt} = 
\frac{|\lambda^{dud}_a|^2}{96\pi s^2M^4}\!
\left[4s^2 +
4s(t-m_{X_a}^2) + t(t-m_{X_a}^2)\right]
\label{operator-2}
\ee
hereafter $s\equiv(p_1+p_2)^2$, $t\equiv(p_1-k_1)^2$.  
\item  Operators ${\cal O}^{dus}$ and ${\cal O}^{dub}$. Here the main
  subprocess is $d(p_1)u(p_2)\rightarrow (\bar{s},\bar{b})(k_1)
  X_{a}(k_2)$ with differential cross section
\be
\frac{d\hat{\sigma}}{dt}=\frac{|\lambda^{dus(b)}_a|^2}{96\pi s^2M^4}
(s+t-m_{X_a}^2)(s+t) 
\label{operator-3}
\ee
\item Operator ${\cal O}^{dtd}$. As in the previous case here we
  consider the dominant subprocess  $d(p_1)d(p_2)\rightarrow
  \bar{t}(k_1) X_{a}(k_2)$ whose differential cross section reads
\be
\begin{split}
\frac{d \hat{\sigma} }{dt}=\frac{|\lambda^{dtd}_a|^2}{96\pi s^2M^4}
\big[&4t^2\!+4t(s\!-\!m_{X_a}^2\!\!-m_t^2)\\ 
&+
  s(s\!-m_{X_a}^2\!\!-m_t^2)+4m_{X_a}^2m_t^2\big],  
\end{split}
\label{operator-4}
\ee
where we take into account the top quark mass.
\end{itemize}

For integration over the phase space with a particular set of cuts we
use CompHEP package\,\cite{CompHEP} with
CTEQ6L1\,\cite{Pumplin:2002vw} as a universal PDF set. The
interactions of the type like the first term in Eq.~\eqref{eq:1} can
not be introduced in the CompHEP directly (this soft can not deal with
color singlet operators like
$\epsilon_{\alpha\beta\gamma}q^{\alpha}q^{\beta}q^{\gamma}$). We
generate some analog of considered process and {\it replace} the
matrix element squared with those directly calculated
from~\eqref{first-a}-\eqref{second-b} (we have checked 
that the results \eqref{operator-1}-\eqref{operator-4} are reproduced
by the code).

To obtain exclusion plots for model parameters we use the latest
results of CMS searches for jet$+E_T^{\rm
miss}$~\cite{Khachatryan:2014rra}. These  searches at $\sqrt{s}=8$ TeV
with 19.7 fb$^{-1}$ integrated luminosity adopted cut on jet rapidity
$|\eta_{jet}|<2.4$  
and a cut on transverse missing energy\,\footnote{Moreover, 
the event selection procedure of~\cite{Khachatryan:2014rra} allows a
second jet which is separated from the first by less that 2.5 azimuth
radian, $\Delta \phi(j_1,j_2)<2.5 $. Such angular requirement
suppresses dijet QCD events. 
}.
The results of the
analysis are presented as upper limits on number of events for a given
cut on $E^{miss}_T$. We have found that the most stringent bounds on
parameters of the model in question are obtained for the cut
$E_T^{miss}>500$~GeV. Note that $E_T^{miss}$
in~~\cite{Khachatryan:2014rra} is defined as the magnitude of vector
sum of transverse momentum of all particles in the event. For our
calculations this means that $E_T^{miss}$ coincides with transverse
momentum of jet, $p_T^j$.

Using the procedure described above we calculate the 
number of signal events expected for some set of parameters and
compare it with an observed 95 $\%$ CL upper limit on number of events 
from new physics $N^{UL}_{obs} = 120$ for a given luminosity (see
Table~3 in Ref.~\cite{Khachatryan:2014rra}).  Note that we use tree
level expressions for cross sections and expect that QCD corrections 
somewhat increase the values.  So, the bounds obtained
below are only conservative.

\begin{figure}[!htb]
\centerline{\includegraphics[width=0.49\columnwidth]{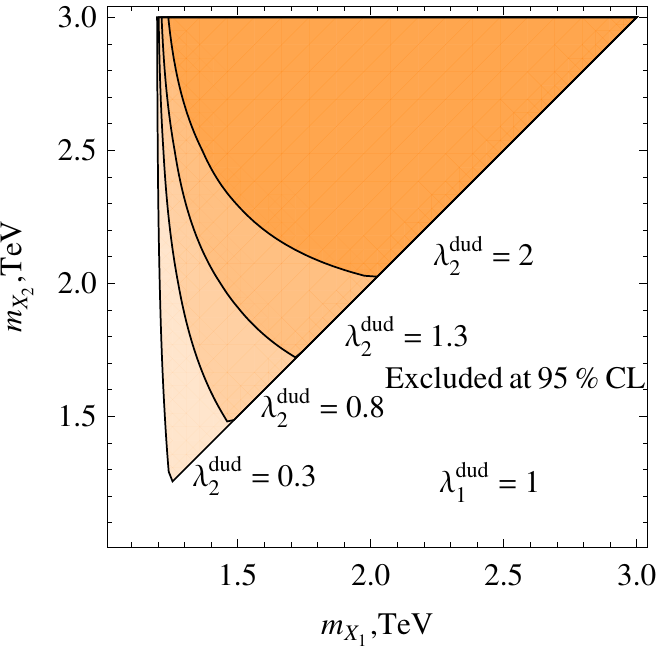} 
\hskip 0.02\columnwidth
\includegraphics[width=0.49\columnwidth]{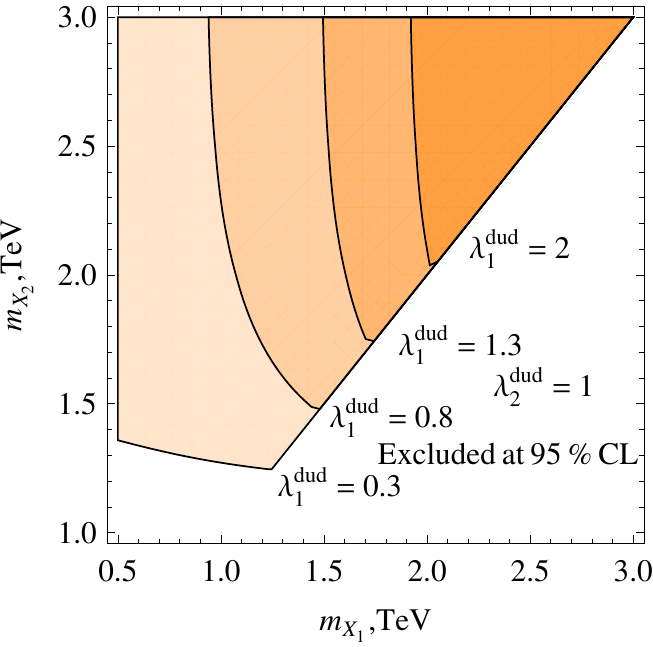}}
\caption {{\it Left panel}: allowed regions in $(m_{X_1},
  m_{X_2})$ plane  for different values of
  $\lambda^{dud}_2$ and  
  $\lambda^{dud}_1=1$. {\it  Right panel}: allowed regions in $(m_{X_1},
  m_{X_2}) $ plane for different values of $\lambda^{dud}_1$ and 
  $\lambda^{dud}_2=1$. We take $M=3.5$\,TeV and $m_{X_2}> m_{X_{1}}$.
\label{fig:1}} 
\end{figure}

For a given operator of the type\,\eqref{eq:1} 
the relevant model parameters are scale $M$, masses of heavy
mediators $m_{X_1}, m_{X_2}$ and coupling constants $\lambda_1,
\lambda_2$. To illustrate the obtained bounds we fix 
some of these parameters. We start with operator ${\cal
  O}^{dud}$. 
In Fig.\,\ref{fig:1} 
\begin{figure}[!htb]
\centerline{\includegraphics[width=0.49\columnwidth]{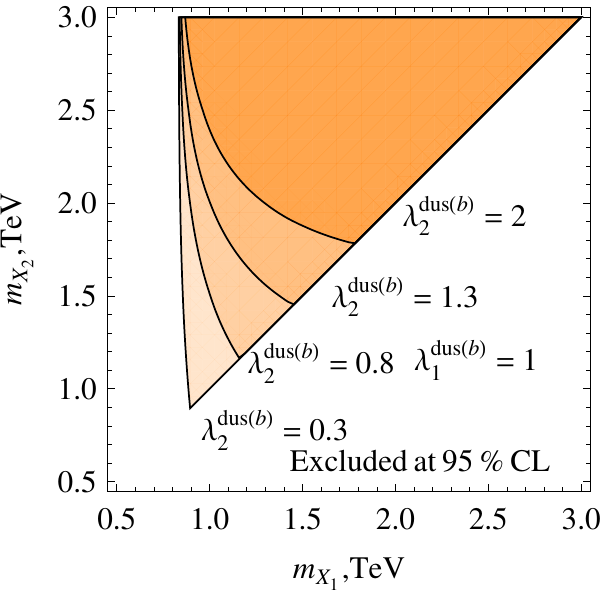} 
\hskip 0.02\columnwidth
\includegraphics[width=0.49\columnwidth]{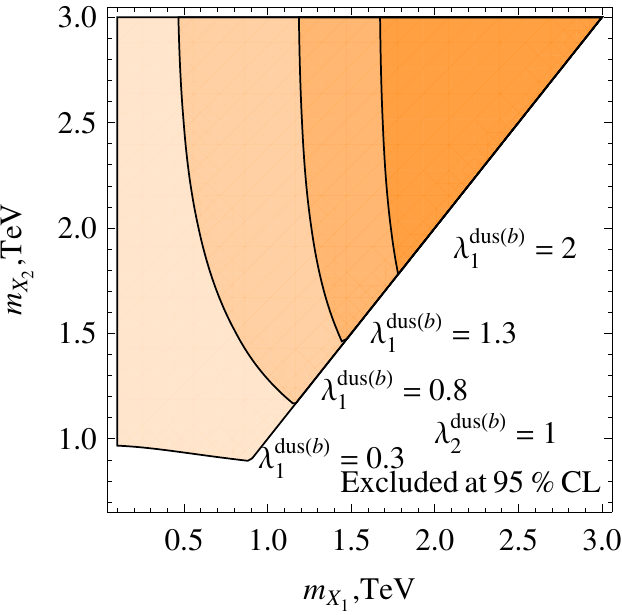}}
\caption {{\it Left panel}: allowed regions in $(m_{X_1}, m_{X_2})$ plane
  for different values of $\lambda^{dus(b)}_2$ and 
  $\lambda^{dus(b)}_1\!=\!1$. {\it  Right panel}: allowed regions in $(m_{X_1},
  m_{X_2}) $ plane for different values of $\lambda^{dus(b)}_1$ and 
  $\lambda^{dus(b)}_2=1$. We take $M=2.5$\,TeV and $m_{X_2}>
  m_{X_{1}}$.
\label{fig:7}} 
\end{figure} 
we show allowed regions in $(m_{X_1}, m_{X_2})$
plane for $M=3.5$\,TeV. On the left plot we fix 
$\lambda^{dud}_1=1$ and show the allowed regions varying $\lambda^{dud}_2$
from 0.3 to 2. 
We see that the limit on the mass
of $X_1$ varies from 1.2 to about 2\,TeV depending on $\lambda^{dud}_2$. 
The limit is sensitive to $\lambda^{dud}_2$ in the
region of light $X_2$. On the right plot we fix 
$\lambda^{dud}_2=1$ and show the allowed regions
for different values of $\lambda^{dud}_1$ as in the previous case. The
obtained limits on masses of $m_{X_1}$ and $m_{X_2}$ in this case are 
up to 1.9\,TeV for $m_{X_1}$ and in the interval
$1.2-2.0$~TeV for $m_{X_2}$. Note that the hierarchy 
$m_{X,1}\ll m_{X,2}$ although allowed by collider searches is
in some tension with requirement of successful baryogenesis as
illustrated by Eq.~\eqref{BAU} and subsequent discussion in
Sec.~\ref{sec:1}.

Events with single $b$- or $s$-jets are equivalent from view
point of models (\ref{first-b}) and (\ref{second-a}). In
Fig.\,\ref{fig:7} we show allowed regions in $(m_{X_1}, m_{X_2})$
plane for both ${\cal O}^{dus}$ and ${\cal O}^{dub}$ interactions
(which are  practically the same due to smallness of $m_b$) for
$M=2.5$ TeV.  Note, that bounds of the same order are expected for 
operators~\eqref{first-a}-\eqref{second-a} with different permutations
of the quark fields.  

\begin{figure}[!htb]
\includegraphics[width=\columnwidth]{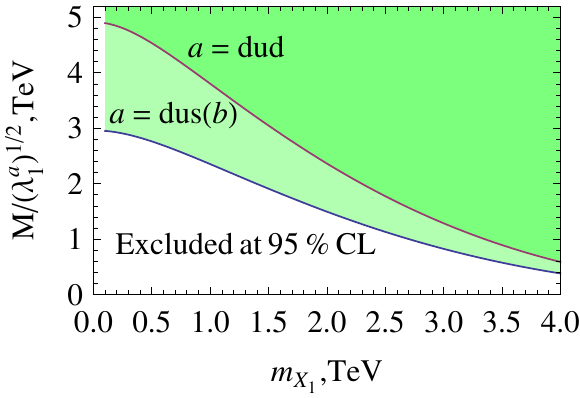}
\caption{
Exclusion region in $(M/\sqrt{\lambda^{a}_1}, m_{X_1})$ plane 
for $\lambda^{a}_2=0$.
 \label{fig:2a}}
\end{figure}

Next, we can consider a limiting case when $m_{X_2}\gg m_{X_1}$ and
neglect contribution to the cross section from the heaviest mediator.
Equivalently, here we set $\lambda^{dus(b)}_2=\lambda^{dud}_2=0$. In
this case we obtain an exclusion plot in $(m_{X_1},
M/\sqrt{\lambda^{a}_1})$ plane shown in Fig.\,\ref{fig:2a}. Here $a$
labels the type of interaction, $a=dus(b),dud$. 

From this plot we can see how the limit on $m_{X_1}$ scales 
with combination $M/\sqrt{\lambda^{a}_1}$ in front of
the interaction lagrangian \eqref{eq:1}. We note that for small values of
$M/\sqrt{\lambda^{a}_1}$ (i.e. on the right part of this exclusion
plot) formally the limits on mass of $X_1$ become stronger. In this case
the effective theory \eqref{eq:1} with the interaction term 
suppressed by $M$ may be invalid and instead a full
theory should be considered. But in this case the actual bounds will
depend on UV completion of the model and in particular the width of
$X_a$ resonances, see e.g. \cite{Blinov:2012hq}.   

It follows from (\ref{operator-1}), (\ref{operator-2}) and
(\ref{operator-3}), that the rate of the process $pp\rightarrow
X_1 \bar{s}(\bar{b})$ is less than the rate of $pp\rightarrow X_1 \bar{u}$,
so that the lower bounds on parameter
$M/\sqrt{\lambda_{1}^{dus(b)}}$ are below the limits on 
$M/\sqrt{\lambda_{1}^{dud}}$. The exclusion region for operator ${\cal 
O}^{dud}$ is wider than that for ${\cal O}^{dus}$.

\section{Further tests at LHC}
\label{sec:3}

\begin{figure}[!htb]  
\centerline{\includegraphics[width=0.5\columnwidth]{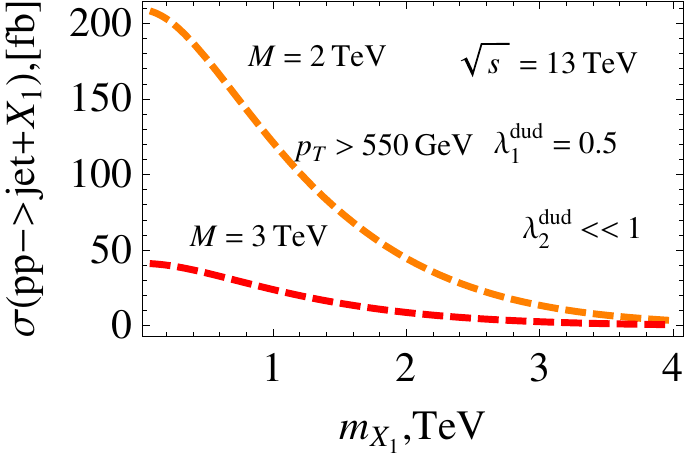} 
\includegraphics[width=0.5\columnwidth]{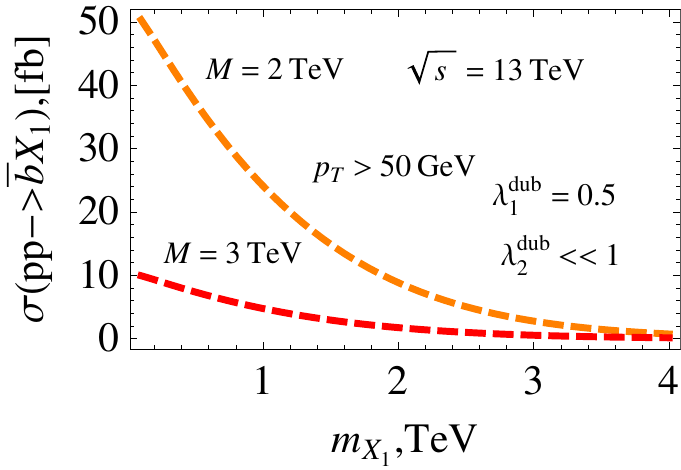}}
\caption{
{\it Left panel}:  cross section of process $pp\to
\bar{d}X$ at $\sqrt{s}=13$\,TeV. {\it Right panel:} cross section of
process $pp\to \bar{b}X$ at $\sqrt{s}=13$\,TeV. 
 \label{fig:2b}}
\end{figure}

\begin{figure}[!htb]
\centerline{\includegraphics[width=0.5\columnwidth]{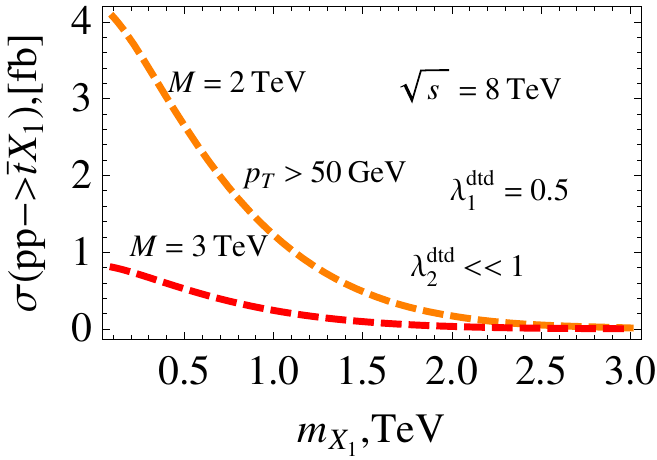} 
\includegraphics[width=0.5\columnwidth]{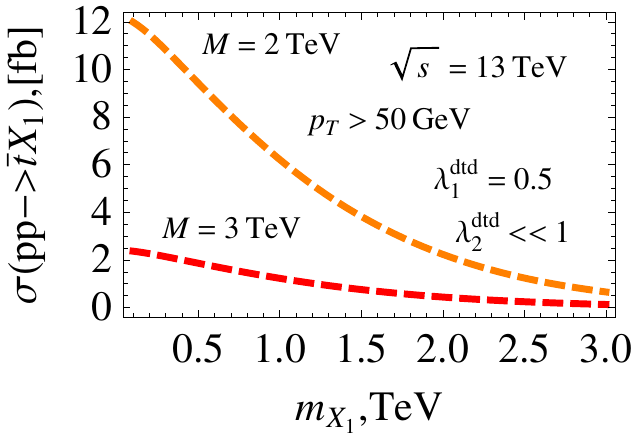}}
\caption {{\it Left panel}: cross section for the process $pp\to
\bar{t}X$ at $\sqrt{s}=8$~TeV. {\it Right panel}: cross section for the process
$pp\to \bar{t}X$ for $\sqrt{s}=13$~TeV. 
 \label{fig:3}}
\end{figure}

Here we calculate tree level cross sections for processes 
\be
pp\to\bar{d}X \qquad
pp \to\bar{b}X,  \qquad pp\to \bar{t}X
\ee
at $\sqrt{s}=13$~TeV. Note that the two last processes involve
operators which can not be probed by IND process. In Fig.~\ref{fig:2b}
we plot the cross section for $pp\to \bar{d}X$ (right panel) and
$pp\to \bar{b}X$ (left panel) at $\sqrt{s}=13$~TeV.  
Here we fix $\lambda^{dud} (\lambda^{bud}_1)=0.5$ and neglect
contribution from $X_2$.  Integration over the phase space is
performed with the following cuts: $|\eta|< 2.4$ and $p_T > 550$~GeV
for $pp\to \bar{d}X$ and $p_{T} > 50$~GeV for
$pp\to \bar{b}X$. Obtained cross section is up to 200~fb for
$\bar{d}X$ final state and of order 10-50\,fb for $\bar{b}X$.  
In Fig.\,\ref{fig:3} 
we plot the cross section of process $pp\to \bar{t}X$ for
$\sqrt{s}=8$\,TeV (left panel) and $\sqrt{s}=13$\,TeV (right panel) 
for comparison. The process has monotop signature (see
e.g~\cite{Agram:2013wda}). The current limits on monotop searches from 
results of ATLAS~\cite{Khachatryan:2014uma} and CMS~\cite{Aad:2014wza}
are of order $0.1-1$~pb and the limits on parameters of the
hylogenesis model are weak. 

We finish the Section with brief discussion of other rather specific
signatures. The successful baryogenesis requires 
sufficient CP-asymmetry in decays
$X_1\to udd$ and $\bar{X}_1\to\bar{u}\bar{d}\bar{d}$. The expression
for partial width of these decays is given by~(\ref{B}).
Comparing (\ref{B}) with width of the dominant decay (\ref{A})
one can see that for $|\lambda_1|\sim |\zeta_1|\sim 1$ and for
$m_{X_1}\lesssim M$ the branching of $X_1$ decay to quarks can
reach values $5\times 10^{-3}$. Even larger values are possible if
$|\zeta| < |\lambda|$. So, the invisible decay of $X_1$ is not the
only signature to search for hylogenesis at LHC. Another interesting
possibility appears when produced $X_1$ decays into three quarks and
we have four-jet signature. For ${\cal O}^{dud}$ operator 
processes of interest are 
\be 
ud\rightarrow \bar{d}X_{1} \rightarrow \bar{d} \,udd, \qquad
dd\rightarrow \bar{u}X_{1} \rightarrow \bar{u}\,udd.
 \label{pp_4j}
\ee
In case of ${\cal O}^{dub}$ operator the relevant to searches for
 hylogenesis signals are  
 4-jets (involving 2b-jets) in the final state, 
\be
 ud\rightarrow \bar{b} X_{1}\rightarrow  \bar{b}\, dub.
 \label{ud_bdub}
\ee
For ${\cal O}^{dtd}$ events with 2jets$+t\bar{t}$ arise
from subprocess
\be 
dd\rightarrow \bar{t}X_{1} \rightarrow \bar{t}\,tdd.
 \label{pp_2jtt}
\ee
 We leave an analysis of  these signatures
for future study.



\section{Conclusions}
In this paper we explore collider phenomenology of the hylogenesis
model. We have found that the current searches for jet and missing
transverse energy signature at ATLAS and CMS place bounds on mass
of the lightest mediator $X_1$ of order $0.7-2$~TeV 
for interactions mediated by operators ${\cal
O}^{dud}$, ${\cal O}^{dus}$ and ${\cal O}^{dub}$ suppressed by scale
$M=2.5-3.5$~TeV. At the same time
current monotop searches do not allow to put considerable bounds on
the parameters of the model for $M$ in TeV range. For the next run of
LHC  experiments at $\sqrt{s}=13$~TeV we predict cross section up to
200~fb for jet$+E_T^{miss}$ searches and up to several tens fb for monotop 
process. Four jets signature (including 2jets+$b\bar b$ and
2jets+$t\bar t$) is suggested to probe hylogenesis at LHC. 

\paragraph*{Acknowledgments}
The work was supported by the RSCF grant 14-12-01430.


\end{document}